\documentclass[%
reprint,
superscriptaddress,
floatfix,
amsmath,
amssymb,
aps,
notitlepage
]{revtex4-1}

\usepackage{graphicx}
\usepackage{mathrsfs}
\usepackage{tikz,pgfplots}
\usepackage{bm}
\usepackage[hidelinks=true]{hyperref}
\usepackage{xcolor}

\definecolor{greenish}{rgb}{0.13,0.58,0.16}
\definecolor{reddish}{RGB}{174,12,48}
\definecolor{blueish}{rgb}{0.12, 0.56, 1.0}
\definecolor{magenta}{rgb}{0.8, 0.0, 0.8}

\hypersetup{
colorlinks   = true, 
urlcolor     = greenish, 
linkcolor    = magenta, 
citecolor   = blueish 
}

\def\d{\textrm{d}}
\def\dvec{\hat{\bm{d}}}

\def\f12{\frac{1}{2}}

\def\N{\bm{N}}

\def\m{{\rm m}}

\def\ga{\gamma}

\def\x{\bm{x}}

\def\k{\varkappa}
\def\g{\gamma}

\def\hk{\tilde{\varkappa}}
\def\hs{\tilde{s}}

\def\lec{l_{\rm bc}}

\def\O{\mathcal{O}}
\def\bb{\Omega^{-1}}
\def\ibb{\Omega}

\def\x{\mathbf{x}}
\def\hz{\hat{\bm{z}}}

\def\ga{\gamma}

\def\ec{{\rm bc}}
\def\eg{{\rm eg}}

\def\nm{\textrm{nm}}
\def\mm{{\rm mm}}
\def\cm{{\rm cm}}
\def\S{\bm{\mathcal{S}}}
\def\hS{\tilde{\bm{\mathcal{S}}}}
\def\N{\bm{\hat{N}}}

\def\Nd{\N \times \dvec_3}
\def\E{\mathscr{E}}

\def\hE{\tilde{\mathscr{E}}}

\def\vkap{\bm{\varkappa}}
\def\um{\mu \text{m}}
\def\varphi{\Phi}

\def\htl{\hat{\bm{t}}}
\def\hx{\hat{\mathbf{x}}}
\def\MPa{\text{MPa}}

\newcommand{\blue}[1]{\textcolor{black}{#1}}

\begin{document}
	\title{Shapes of a filament on the surface of a bubble}
	\author{S Ganga Prasath}
	\email{gangaprasath@seas.harvard.edu}
	\affiliation{School of Engineering and Applied Sciences, Harvard University, Cambridge MA 02138, USA..}
	\affiliation{International Centre for Theoretical Sciences (ICTS-TIFR) Shivakote,\\ Hesaraghatta Hobli, Bengaluru 560089, India.}
	\author{Joel Marthelot}
	\affiliation{Aix-Marseille University, CNRS, IUSTI (Institut Universitaire des Syst\'emes Thermiques Industriels), 13013 Marseille, France.}
	\author{Rama Govindarajan}
	\affiliation{International Centre for Theoretical Sciences (ICTS-TIFR) Shivakote,\\ Hesaraghatta Hobli, Bengaluru 560089, India.}
	\author{Narayanan Menon}
	\affiliation{Department of Physics, University of Massachusetts Amherst, Amherst, MA 01003, USA.}

	\date{}

	\begin{abstract}
The shape assumed by a slender elastic structure is a function both of the geometry of the space in which it exists and the forces it experiences. We explore by experiments and theoretical analysis, the morphological phase-space of a filament confined to the surface of a spherical bubble. The morphology is controlled by varying bending stiffness and weight of the filament, and its length relative to the bubble radius. When the dominant considerations are geometry of confinement and elastic energy, the filament lies along a geodesic and when gravitational energy becomes significant, a bifurcation occurs, with a part of the filament occupying a longitude and the rest along a curve approximated by a latitude. Far beyond the transition, when the filament is much longer than the diameter, it coils around the selected latitudinal region. A simple model with filament shape as a composite of two arcs captures the transition well and for better quantitative agreement with the subcritical nature of bifurcation, we study the morphology by numerical energy minimization. Our analysis of filament's morphological space spanned by a geometric parameter, and one that compares elastic energy with body forces, may provide guidance for packing slender structures on complex surfaces.
    \end{abstract}

	\maketitle
	
\section*{Introduction}
The spatial conformation of a slender elastic structure is a function both of the geometry of the space in which it exists as well as the forces it experiences. An unconfined elastic filament does not bend or stretch, but when the elastic filament is confined to a two dimensional surface, the geometrical properties of the surface plays a role in determining its morphology~\cite{huynen2016surface, guven2014environmental}. Such surface confinement can arise in a variety of situations such as a DNA wrapping around histone~\cite{swigon1998elastic, tobias2000elastic, coleman2000elastic}, plants climbing along complex topographies~\cite{goriely2006mechanics}, packaging of optic fibre cables around curved surfaces~\cite{seemann1996deformation} to long drill strings beneath the earth's surface~\cite{van2001static, tan1993buckling,wu1993helical1,wu1993helical2}. The role of surface geometry in determining the shape is crucial in all of these situations, however in each example the filaments are also subject to forces (electrostatic, active internal stresses, gravity, friction) that affect their conformation. For example a filament with intrinsic curvature held at one of its ends, such as a hair~\cite{miller2014shapes}, takes shapes determined by the competition between its own weight and the bending stiffness of the structure~\cite{mahadevan1996coiling, mahadevan1999periodic}, where for small mass density the filament retains its natural shape, and for denser filaments, the structure undergoes a shape transition to a morphology determined by the weight.

In this article we explore the general question of how substrate geometry and external force combine to determine the shape of a filament by considering the specific case of an elastic filament on the surface of a spherical bubble. When a pre-wet filament is introduced on the surface of a droplet or a bubble, if the droplet's Laplace pressure exceeds a critical value, the  filament buckles  ~\cite{elettro2016,bico2004elastocapillary,prasath2021,brau2018,schulman2017elastocapillary}. In the buckled state the shape of the drop can change in order to minimize the surface energy of the droplet and the elastic energy of the filament~\cite{py2007,prasath2021}. However, since we wish to explore the mechanics of the filament in a fixed confining geometry, we work with highly-bendable filaments so that the spherical geometry of the substrate remains unchanged by the filament. In this high-bendability limit, we perform experiments where the forces due to gravity and bending compete to determine the morphology of a filament confined to the surface of a spherical pendant bubble. When the gravitational potential energy of the filament is negligible compared to its bending energy, we find that the filament lives along a geodesic -- a longitude of the sphere -- where it adopts the radius of curvature of the bubble.  However when the effects of gravity become important, the filament undergoes a transition into a complex shape. Beyond this transition, the filament lies partly along a longitude and partly along a latitude at a particular polar angle.  A calculation which accounts for the  energy of these two segments of the filament correctly captures the shape below the transition length as well as the threshold control parameters for bifurcation from the geodesic shape. However, the approximate theory fails to capture the quantitative details of the shape after the transition. A numerical minimization of the competing energies in the system identifies deviations from the approximate theory and captures accurately the shapes in experiments. On further increase in the length of the filament, the filament enters a coiled phase with the filament aligned on top of itself in layers. In what follows, we explore experimentally and by numerical simulation, a two-parameter phase diagram of the morphology of the filament in a curved space: one parameter is geometric, and describes the length of the filament versus the radius of confinement; the other parameter encapsulates the competition between elasticity and the external forces (gravity, in this case).

\begin{figure*}[ht!]
	\centering
	\includegraphics[width=0.75\textwidth]{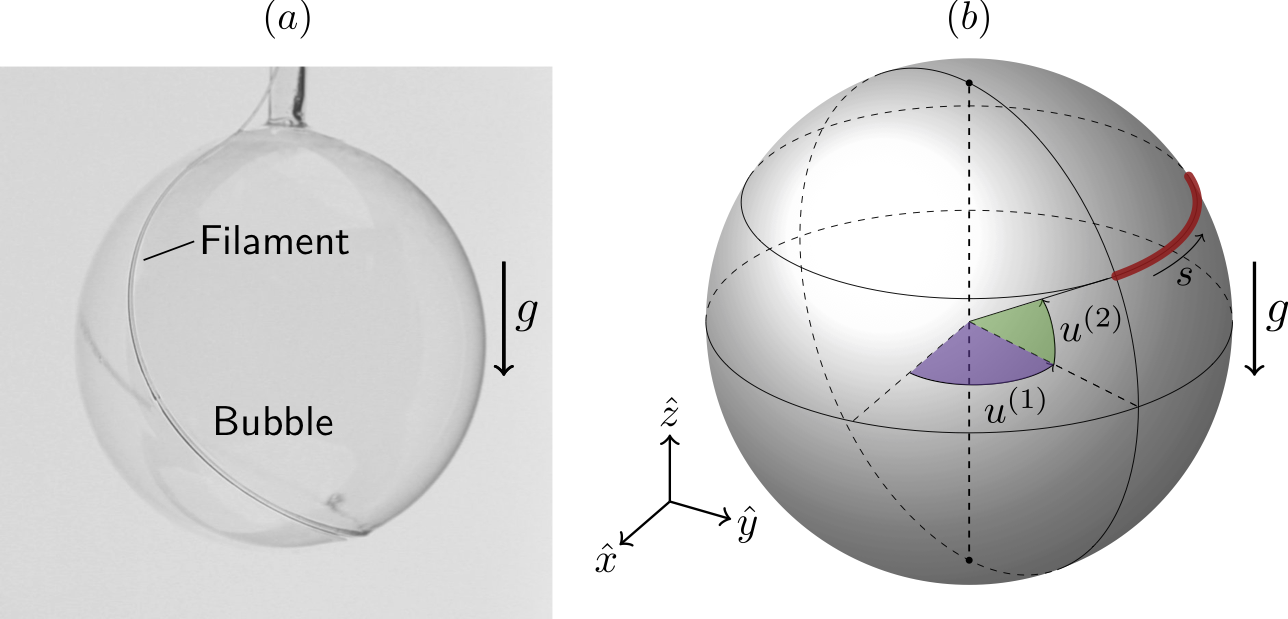}
	\caption{$(a)$ A soap bubble hanging from the end of a capillary tube of diameter $1 \mm$ with gravity pointing downwards. The elastic filament sits on top of the bubble with one end hinged to the capillary tube. $(b)$ Schematic of the setup in which the bubble is approximated by a sphere with $u^{(1)}$ being the azimuthal angle, and $u^{(2)}$ being the polar angle on the sphere. Gravity points downwards, towards the negative $z$-axis. The filament shown in red is represented using $( u^{(1)}(s), u^{(2)}(s) )$ where $s$ is the arc-length along the filament.}
	\label{fig:phase}
\end{figure*}

\section*{Experiments and length-scales}
In our experiments the spherical surface is provided by a pendant soap bubble hanging at the end of a capillary tube as shown in Fig.~\ref{fig:phase}$(a)$. The bubble is made using a liquid soap (commercial DAWN$^{\rm TM}$) in a 1:4 water-glycerol solution. The high concentration of glycerol with viscosity 1000 cSt helps reduce the drainage rate providing longer bubble survival time. We then place a thin elastic filament made out of silicone on top of the bubble. These filaments are made by first melting a rod of silicone at $300^\circ$C using a silicone glue gun (\blue{Aptech Crown}) and then pulling a small droplet of the melt with tweezers which then sets in a few seconds. The diameters of the filaments made by this procedure can be varied between $\O(10 \um) - \O(100\um)$ by varying the pulling rate manually. The diameter is measured under an optical microscope, where we also confirm that we have a filament of uniform thickness. We also soak the filaments in Sudan Red G dye (SIGMA 17373) solution and dry them to make them visible in white light.  The bubble is illuminated by a diffuse background light and we use a Nikon DLSR 5000 camera to observe the conformation of the filament on the bubble. We wet the filaments in the soap solution before placing them on the bubble surface in order to eliminate the effects of the filament surface energy on the filament morphology. \blue{This procedure also helps us maintain longer bubble survival times as a dry filament can cause sudden changes in the surface energy of the interface resulting in rupture. When the wet filament is placed on the bubble, the change in the surface energy of the bubble is only due to distortion of the fluid film around the thickness of the filament.} In the experiments one end of the filament is held at the end of the capillary tube and the other end is left free. The hinged end of the filament experiences no torque and can rotate freely about the end of the capillary tube. For a given filament of length $L$, we quasistatically increase or decrease the bubble radius $R_b$ using a microfluidic pump attached to the other end of the capillary tube.

\begin{figure*}
	\centering
	\includegraphics[width=0.75\textwidth]{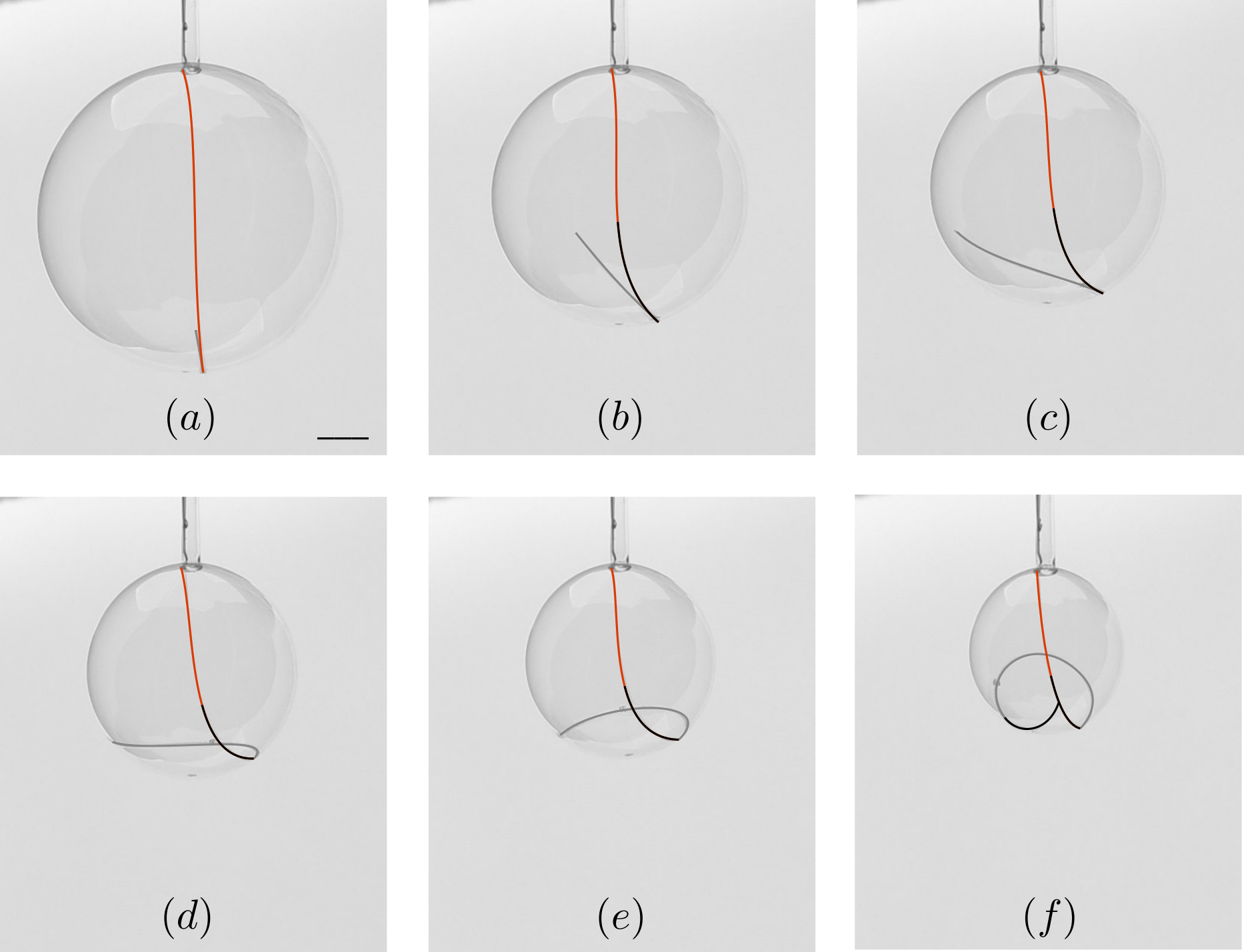}
	\caption{$(a-f)$ Filament morphology on the surface of the bubble for different bubble sizes for a fixed filament length of $L = 30 \mm$. For large bubble size the filament stays along a longitude and as the bubble size is decreased, assumes more complex shapes where it bends smoothly from the longitude along a chosen latitude. The solid line is traced along the filament for higher contrast, and is shown in orange color along the longitude, and in black where it deviates from a longitude. In $(f)$ we show the filament in self-contact. Coiling occurs on further decrease in bubble size. The scale bar is $3\mm$.}
	\label{fig:seq}
\end{figure*}

In order to describe the phenomenon we observe in our experiments, we collect in this section the relevant length-scales that play a role in the determining the filament morphology. There are several length-scales associated both with the filament and the bubble; we attempt to reduce the complexity by working in regimes where some of these are irrelevant. Since we expect gravity to be relevant in our experiments, we define the elasto-gravity length-scale $l_{\eg}\sim (2 EI/(\varrho g \pi t^2))^{1/3}$, where $E$ is the Young's modulus of the filament material, $I$ the area moment of inertia of the filament, $\varrho$ the material density, $t$ the filament thickness and $g$ the gravitational constant. This is the length scale associated with filament deformation when the weight of the filament is balanced by bending forces. In the limit of strong capillarity, where the bubble does not deform, the radius of the bubble $R_b$ is the only length-scale associated with the 2D spherical surface. Using these two length-scales we can define a non-dimensional number that quantifies the relative importance of gravity and bending as the \textit{elasto-gravity} bendability, $\bb_g=(l_{\eg}/R_b)^3=2 EI/(\varrho g \pi t^2 R_b^3)$. The larger the value of $\ibb_g$, the stronger the effects of gravity and the smaller the $\ibb_g$, the stronger the effects of bending. Another non-dimensional number arising purely out of geometry is $\Phi=(L/R_b)$. We call this non-dimensional number the \textit{coiling parameter}, since as we will see later this number helps describe the state of coiling. The two non-dimensional numbers of interest thus are the coiling parameter $\Phi$ and the elasto-gravity bendability $\bb_g$ and the different morphologies of the elastic filament on the spherical bubble are described in this phase space spanned by these two nondimensional groups. 

In identifying the relevant length-scales in our experiments, we have tacitly made assumptions of scale separation in the filament-bubble system that we detail here. For a spherical bubble made of a fluid of surface tension $\g$, fluid density $\varrho_f$, radius $R_b$ and a filament of length $L$, Young's modulus $E$, material density $\varrho$, and thickness $t$, there are six length-scales that  at play in the experiments. These are $L, R_b, t$, capillary length $l_c \sim \sqrt{\g/\varrho_f g}$, bendo-capillary length $l_{\ec} \sim (EI/\g)^{1/3}$ and finally, the  elasto-gravitational length scale $l_{\eg}$ described above. Since we want the bubble to remain spherical throughout the experiments, we ensure this by keeping the capillary length to be the largest length scale in the system. Moreover we want the surface energy of the bubble to not affect filament bending i.e., we want the bendo-capillary length to be very small such that the filament is able to completely lie on the bubble without affecting the bubble. This is taken care of by operating in a scale-separated regime: $t \ll l_{\ec} \ll (L, R_b, l_{\eg}) \ll l_c$. When the relevant length-scales are $(L, R_b, l_{\eg})$, the filament morphology is governed by the geometry, the bending force and the self-weight of filament. This separation of scales in our experiments is established by choosing a thin ($t \sim \O(100 \mu\m)$) and soft filament ($E = 1\MPa$) with a bubble size much larger than filament diameter ($L, R_b \sim \O(\cm)$) leading to an bendo-capillary length-scale $l_{\ec} \sim \O(\mm)$.
The ratio of capillary force and bending force can be quantified using inverse capillary bendability, $\bb = (l_\ec/R_b)^{3}$. Lastly, in our experiments the tangential strain along the filament is small and thus are in the inextensible \textit{elastica} limit~\cite{audoly2010elasticity}. This is ensured by choosing $l_m = \ga/E \sim \O(\nm) \ll t$. We explore the phase space of filament morphologies by changing the filament length $L$, the bubble radius $R_b$ and the filament thickness $t$ while keeping all the other experimental parameters fixed.

\section*{Results}\label{sec:res}
When one end of the filament is hinged to the capillary tube at the north pole of the bubble, with gravity pointing from north to south, the filament takes different morphologies, as shown in Fig.~\ref{fig:seq}, when the ratio $\Phi$ of filament length to bubble radius is increased. Initially, the filament  lives along a  longitude (see Fig.~\ref{fig:seq}$(a)$). For small filament lengths we expect gravity to not play a role in determining the filament configuration, so the shape is determined purely by minimizing the bending energy. For a sphere the great circle has the smallest curvature and thus the filament lies along a longitude, since bending energy $\E \sim \varkappa^2 L$, $\varkappa$ being the curvature. As we increase the ratio $\Phi$, the conformation of the filament changes from that of a longitude, as shown in Fig.~\ref{fig:seq}$(b-d)$.  Intuitively we recognise that this effect is due to gravity: the weight of the filament pulls it towards the bottom of the sphere.  At the largest values of $\Phi$, the filament coils around a fixed latitude in the bottom half of the sphere. 

\begin{figure*}
	\centering
	\includegraphics[width=0.85\textwidth]{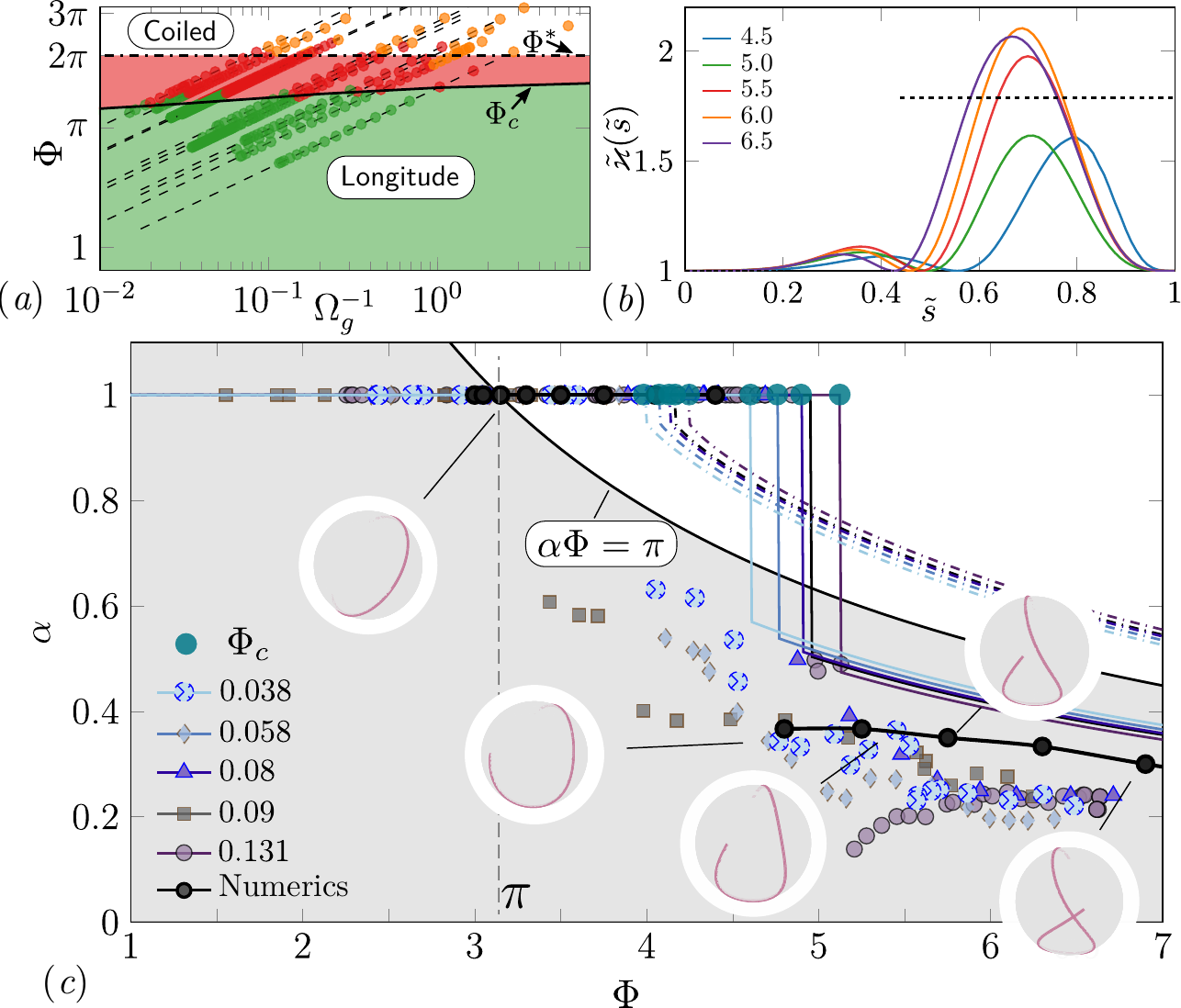}
	\caption{$(a)$ Phase diagram of coiling represented by the non-dimensional elasto-gravity bendability, $\bb_g$ and the coiling parameter, $\Phi$. Circles are experimental data points for a given combination of $ \Phi, \bb_g$. The green region is when the filament stays along the longitude, separated from the red region by the solid line representing the theoretical critical coiling parameter $\Phi_c(\bb_g)$, which represents the regime of deviations from longitude to the more complex shapes seen in Fig.~\ref{fig:seq}. \blue{$\Phi^*$ is the coiling parameter at which coiling starts for $\bb_g \rightarrow + \infty$ and we see that $\Phi_c \rightarrow \Phi^*$ for large $\bb_g$}. The white region, with orange experimental points is the regime of coiling. $(b)$ Curvature along the filament, $\hk(\hs)$ as a function of arc-length, $\hs$ from numerical solutions for different values of $\Phi$ (shown in the legend) for a fixed $\bb_g=0.1$. $(c)$ Bifurcation diagram of coiling: Fraction of filament length along longitude, $\alpha$ as a function of the coiling parameter for fixed values of $\bb_g$. Symbols are from experiments and solid lines (dashed lines) from theory by minimizing energy expression in Eq.~\ref{eq:enerFix} with (without) the constraint of $\alpha \Phi < \pi$ for different $\bb_g$ values shown in the inset. The data points in the gray region indicate that the length along the longitude never crosses the south pole. Numerical minimization shown in black for $\bb_g = 0.1$ captures the experimental observation accurately. \blue{We also show the critical coiling parameter from the model, $\Phi_c$, in turquoise.} Inset also shows different shapes we obtain from numerics for different filament lengths.
	} \label{fig:alpha}
\end{figure*}

We explore the phase space of morphologies, spanned by the two variables $\Phi$ and $\bb_g$, to ensure that the physical mechanism we describe is indeed what we observe in experiments. For a fixed length of the filament hinged at the north pole a change in the radius of the bubble $R_b$ changes both variables, causing us to traverse the morphological phase-space along trajectories $\Phi \sim 1/R_b$ and $\bb_g \sim 1/R_b^3$. We perform these experiments with eleven different filament lengths $L=0.5 \cm - 5 \cm$ and two different filament thicknesses $t=70 \mu \m, 100\mu \m$. We show these trajectories for different filament lengths in Fig.~\ref{fig:alpha}$(a)$ by dashed lines. These trajectories and configurations can be reversed by re-inflating the bubble, except in the regime where the filament coils and self-contact of the filament leads to irreversibility in configurations. Along a given trajectory the green dots correspond to a filament configuration along the longitude, and red dots correspond to deviations from the longitude to more complex shapes. Since changing the thickness of the filament changes the bending stiffness ($EI \sim t^4$) and changing bubble size changes the elasto-gravity bendability ($\bb_g \sim 1/R_b^{3}$), we are able to span two orders of $\bb_g$ in the experiments.

The experimental observations are qualitatively explained by our arguments regarding the role of the geometry ($\Phi$) and gravitational potential energy versus elastic energy ($\ibb_g$). However, a quantitative description of the nature of the transition from the longitude is desired. In such a description, we would also like to predict the polar position at which the filament chooses to situate itself  beyond the transition point from a longitude to a complex shape. In order to gain a quantitative understanding we approximate the shape of the filament as an elastic curve on a sphere and calculate the transitions between morphologies  by minimizing the total energy due to bending deformations as well as gravity. We validate this with the experiments and gain further understanding by performing a numerical minimization of the total energy by approximating the filament as a combination of discrete elastic rods.

\subsection*{Geometry and mechanics of the filament}
We describe the morphology of the filament in our experiments using the elastic energy and the gravitational potential energy of the filament. Since stretching and twisting are extremely high-energy deformations, we consider only bending energy in the inextensible limit. The total energy of the filament in this approximation is
\begin{align}
\E = \frac{EI}{2} \int_0^L \varkappa^2(s) \ \d s +  \varrho g \frac{\pi t^2}{4} \int_0^L (\S(s) - \S_o)\cdot \hz \ \d s,
\end{align}
where $\S(s) \equiv \S(u^{(2)}(s),u^{(1)}(s))$ is the location of the filament center-line on the surface of the sphere. $\S_o$ is a reference location about which the potential energy is measured, $\varkappa(s)$ the magnitude of curvature along the arc-length $s$ and $\hz$ the direction against gravity (see Fig.~\ref{fig:phase}$(b)$). For the specific case of a sphere we can explicitly write the parameterization of the surface as:
\[
\S(u^{(2)},u^{(1)}) = w(\cos u^{(2)}  \cos u^{(1)}, \cos u^{(2)} \sin u^{(1)}, \sin u^{(2)}).
\]
The curvature vector $\vkap(s)$ can be decomposed in the orthogonal Darboux frame as:
\begin{align}
\vkap (s) =& \ \k_n \N + \k_g (\Nd),
\end{align}
$\k_n$ being the normal curvature along the surface normal given by
\[
\N = \frac{\S_{u^{(1)}} \times \S_{u^{(2)}}}{||\S_{u^{(1)}} \times \S_{u^{(2)}} ||}.
\]
where $\S_{u^{(j)}}$ denotes derivative with respect to $u^{(j)}, j = 1, 2$ and $\k_g$ is the geodesic curvature along the binormal direction $(\Nd)$, with $\dvec_3$ being the tangent vector along the curve $\S(s)$. The arc-length and the curvature are non-dimensionalized using the bubble radius $R_b$ and the total energy is non-dimensionalized using $(EI/2R_b)$ as the energy scale: $\hs = s/R_b, \hk = \k R_b$ and $\hE = \E / (EI/2R_b)$. 
For a sphere we know that the normal curvature is a constant everywhere, $\k_n = (1/R_b)$ and using this we can rewrite the energy expression as:
\begin{align}
\hE &= \int_0^\Phi [\hk_g^2 + \ibb_g (\hS(\hs) - \hS_o) \cdot \hz] \ \d \hs + \Phi. \label{eq:ener}
\end{align}
where tildes denote non-dimensional variables.

\subsection*{Limit of zero gravity}\label{ssec:zeroG}
In the asymptotic limit of negligible gravity, $\ibb_g \rightarrow 0$, the filament morphology is determined only by the bending energy and the solution amounts to minimizing the bending energy:
\begin{equation}
	\hE \approx \int_0^\Phi \hk_g^2 \ \d \hs + \Phi.
\end{equation}
We know that $\Phi$ remains fixed for a given filament length, $L$ and bubble radius, $R_b$ and thus $\hk_g=0$ or essentially any geodesic is a solution. Since we have hinged one end of the filament, this allows only for longitudes: geodesics that travel from the north pole to the south pole on the bubble. In the weak gravity limit, the boundary condition at the free end, where the filament  experiences a torque $\sim EI/R_b$, does not modify the trajectory from the geodesics (as also pointed out in~\cite{huynen2016surface, guven2014environmental}) which for a sphere are the great circles.  Thus the filament takes this shape for $\Phi \leq 2\pi$. When $\Phi > 2\pi$,  we expect coiling along a geodesic, with deviations from the  geodesic due to self-intersections at finite $t/R_b$. We define the coiling parameter $\varphi^*$ at which coiling starts as the filament length at which self-intersection occurs. For $\bb_g \rightarrow \infty$ the non-dimensional length for coiling is $\Phi^* \rightarrow 2\pi$.

\subsection*{Finite gravito-bendability effects}
In the experiments we have non-zero values for elasto-gravity bendability and as a consequence the filament shape deviates from the geodesic and the critical coiling parameter $\Phi_c$ at which this occurs is expected to be a function of $\bb_g$. The complex shape we see in the experiments beyond the transition from a geodesic can be simplified by approximating it to be a combination of a longitude and a latitude (which is no more a geodesic) at a given polar angle. This approximation ignores the smooth transition between latitude and longitude, treating it geometrically as a sharp kink, with no elastic energy cost. (We evaluate the effects of this approximation later in this article via numerical minimization.) The total energy of such a configuration is given by
\begin{align}
\hE =& \ \hE_l(\alpha \Phi) + \hE_g((1-\alpha) \Phi),
\end{align}
where $\alpha$ is the fraction of the filament along the longitude, $\hE_l$ is the energy due to the filament along the longitude and $\hE_g$ is the energy due to the section along the latitude.
The parameterization of the filament shape lying along any latitude with polar angle $\xi$ is given by,
$u^{(2)} = \xi, u^{(1)} = \hs/\lambda, \hS = (\cos \xi\ \cos(\hs/\lambda), \cos \xi\ \sin(\hs/\lambda), \sin \xi). $
\blue{The inextensibility constraint is enforced by setting $\lambda = \pm \cos \xi$. The energy of this configuration can be calculated to be,
\begin{align}
	\hE_g ((1-\alpha)\Phi;\ibb_g) &=\ (1-\alpha)\Phi\ibb_g (\sin \xi + z_o) \\
	&\ + \int_0^{(1-\alpha)\Phi} \hk_g^2 \ \d \hs + (1-\alpha)\Phi, \nonumber \\
	&=\ (1-\alpha)\Phi\big[ 1+ \tan^2(\xi) + \ibb_g (\sin \xi + z_o) \big] \label{eq:equator}.
\end{align}
We have used the fact that the geodesic curvature of any latitude for a spherical metric as $\hk_g = \tan \xi$~\cite{kreyszig2019differential}}. \blue{Along a longitude the potential energy of any filament configuration depends on where the starting point of the filament $\hs=0$ lies. We can denote an arbitrary longitude by the parameterisation: $ u^{(2)} = \hs-\vartheta, u^{(1)} = 0$ where $\vartheta$ is the polar angle at which the filament is hinged, which we will set to $-\pi/2$ denoting the north-pole. We can now write the potential energy of the configuration as: $\hE_l^p = \ibb_g [\cos \vartheta - \cos (\vartheta-\Phi) + z_o \Phi]$. Since all longitudes are geodesics, $\hk_g=0$ along these curves, leading to the total energy,
\begin{align}
	\hE_{\blue{l}} &= \int_0^{\alpha \Phi} \ibb_g (\x - \x_o)\cdot \hz \ \d \hs + \int_0^{\alpha \Phi} \hk_g^2 \ \d \hs + \blue{\alpha} \Phi, \nonumber \\
	&= \alpha \Phi (1+ \ibb_g z_o) + \ibb_g \big[ \cos \vartheta - \cos (\vartheta-\alpha\Phi) \big].\label{eq:long}
\end{align}
}
Lastly the geometric constraint from inextensibility relates $\alpha$ and $\xi$ as: $\alpha \Phi= (\pi/2 - \xi)$ and after setting $\vartheta=-\pi/2$, the resultant energy can be written as:
\begin{align}
\hE =& \ \alpha \Phi + \ibb_g \sin(\alpha \Phi) + (1 - \alpha) \Phi \big[ \csc^2(\alpha \Phi) + \ibb_g \cos(\alpha \Phi) \big].  \label{eq:enerFix}
\end{align}

The equilibrium configuration is given by extremising the one parameter energy expression~\ref{eq:enerFix}, achieved by solving $\delta \hE/\delta \alpha = 0$ for $\alpha \in [0,1]$. Before we minimize this expression, we must regularize the unphysical divergence in energy $\hE$  for $\alpha \Phi \rightarrow \pi$, which occurs due to the absence of a cutoff length scale associated with capillary bendability. This is done by introducing a cut-off length associated with the capillary bendability (see SI sec.~\ref{sec:reg} for details). The phase boundary for deviation from the geodesic shape is represented by the critical coiling parameter $\Phi_c$ when $\alpha$ first deviates from $1$. We plot this in Fig.~\ref{fig:alpha}$(a)$ as the solid curve and find that our simple geometric theory captures the transition seen in experiments accurately, both at small values of $\bb_g$ as well as $\bb_g \sim \O(1)$. We see that the increase in the value of critical coiling parameter $\Phi_c$ with increase in $\bb_g$ due to reduction in effects of gravity is also captured by the model. Moreover the trend is consistent with the prediction that $\Phi_c \rightarrow \Phi^*$ when $\bb_g \rightarrow \infty$.

\begin{figure*}
	\centering
	\includegraphics[width=0.98\textwidth]{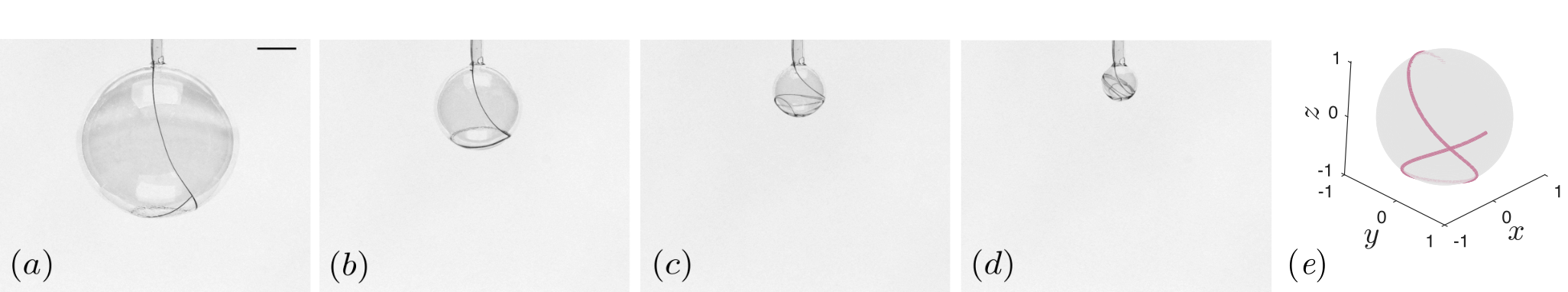}
	\caption{$(a-d)$ Sequence of images in the coiling phase beyond the point of self-intersection where we see the filament starting to pack on the surface of the bubble. For a very small bubble size as in $(d)$ the curvature of the transition zone starts interacting with the hinge at the tip of the capillary tube. Scale bar is $3\mm$. $(e)$ Self-intersecting shape of a filament in numerical simulation beyond the transition threshold for coiling when $\Phi=7$ for $\bb_g=0.1$.} \label{fig:coils}
\end{figure*}

The transition we see in our experiments in Fig.~\ref{fig:seq} from a longitude happens when $\alpha$ changes its value as we tune $\Phi$ for a fixed value of $\bb_g$, however when we change the bubble radius we modify both $\Phi$ as well as $\bb_g$. To compare the value of $\alpha$ from our experiments with the calculation at fixed $\bb_g$, we hold the bubble size fixed while we increase the filament length on the surface by feeding the filament from the top of the bubble. We perform experiments using this protocol for five different values of $\bb_g$ between $0.04-0.1$ and plot the results in Fig.~\ref{fig:alpha}$(c)$. \blue{As the bubble is spherical, we are able to calculate the length of the part of the filament along the longitude from the 3D coordinates of the point where it deviates from a longitude.} In Fig.~\ref{fig:alpha}$(c)$, symbols correspond to experimental measurements, dot-dashed lines for the theoretical minimum energy shapes and solid coloured lines for the minimal energy shape with $\alpha\Phi$ constrained to be less than $\pi$, i.e., the filament cannot cross the south pole. For small $\Phi$, the filament lies perfectly along a longitude, i.e., along $\alpha=1$. This holds true till just beyond $\Phi=\pi$ (dashed vertical line), where the filament is long enough to touch the south pole. At higher $\Phi$, both experiments and theory show deviations from a longitudinal shape, i.e. $\alpha<1$. We emphasize that the theory corresponds to a part of the filament lying perfectly on a longitude and the remainder lying perfectly on a latitude, with a sharp transition between the two segments with no elastic cost. In experiments, however, the transition between these two segments occurs smoothly, and the deformation contributes to elastic energy. Despite this missing energy at the transition zone, the theory predicts the $\Phi_c$ well. However for $\Phi > \Phi_c$, we see deviations in the values of $\alpha$ between the experiments and the theory. \blue{This deviation in magnitude of $\alpha$ is evident in Fig.~\ref{fig:alpha}$(c)$ as all $\alpha$ values from experiments lie below the $\alpha \Phi = \pi$ line while the theoretical predictions of minimal energy without additional constraint on $\alpha$ (dashed colored lines) lie above this line}. But providing an additional constraint of $\alpha \Phi < \pi$ to Eq.~\ref{eq:enerFix} (solid colored lines) results in reasonable agreement with experiments, despite the simplifying assumption made at the transition zone. Since the theory neglects the deformation at the transition zone between the geodesic and the latitude, accounting for this missing length scale would reduce the length of filament on the longitude. This thus would reduce $\alpha$, providing better agreement with experiment without explicitly adding an additional constraint. In order to understand the transition zone, we perform numerical simulations of the filament shapes using the energy in Eq.~\ref{eq:ener}.

\subsection*{Numerical simulation of filament shapes}

We explore the different shapes that the filament takes on the bubble by numerically minimizing the energy in the Eq.~\ref{eq:ener} with the filaments segmented into discrete linked rods. The total energy of the system can be written as
\begin{align}
\hE =& \ \sum_{j} [ \hk_j^2 + \ibb_g z(\hs_j) ] h,
\end{align}
where $\hk_{j} = (\htl(\hs_j) - \htl(\hs_{j+1}))/h$ and $h$ is the length of the discrete rods. Along with this, we further have the constraint that one end of the filament is fixed: $\hx(0) = (0, 0, 1)$ and that all the coordinates of the filament lie along the surface of the sphere: $|| \hx(\hs_j) || = 1$. We use the \texttt{fmincon} function in \textsc{Matlab}'s optimization toolbox to find the minimal energy shape of the filament. We show in Fig.~\ref{fig:alpha}$(c)$ inset the shapes we obtain from the minimization and in Fig.~\ref{fig:alpha}$(b)$ the curvature of the filament for different values of $\Phi$ when $\bb_g = 0.1$. \blue{The  first part of the filament in the numerical solution has constant curvature, corresponding to it lying entirely along a longitude, and beyond some arc length, deviates from this constant curvature. The region deviating from $\hat{\varkappa} \approx 1$ corresponds to the transition zone where effects of the length scale missing in the theory are now accounted for in the numerics.} We compute the value of $\alpha$ as a function of $\Phi$ where we define $\alpha$ as the length from the location where $\hk(\hs)$ reaches its first minimum to the end of the filament. This is shown as black circles in Fig.~\ref{fig:alpha}$(c)$; we find that the bifurcation is indeed sub-critical and for larger values of $\Phi$ we find that $\alpha \sim 1/\Phi$ away from the bifurcation point. The sub-critical nature of the transition can be attributed to a similar mechanism as that seen in the wrapping of a droplet by an elastic sheet (see ref.~\cite{brau2018, kusumaatmaja2011droplet, py2007}) where the external body force that the droplet applies on the sheet makes the flat state unstable and the sheet goes to a wrapped state. The effect of gravity breaks the up-down symmetry of the filament configuration and thus leads to a state where the preferred shape is no more along the longitude. We also look at the bifurcation when the ends of the filament are free in SI sec.~\ref{sec:appdx} and find a similar instability mechanism.

Since the contribution to the total energy of the filament comes from bending and potential energy along the filament latitude and longitude parts as well as the energy of the transition zone, we find the scaling of each of these components. The energy contribution due to bending scales as $\E_l \sim EIl/R_b^2$, gravitation to be $\E_g \sim \varrho g t^2 l^2$ and bending in the transition goes as $\E_{t} \sim EI l_{\eg}(1/R_b^2+1/l_\eg^2)$. These can be written in non-dimensional terms in  with $EI/R_b$ as the energy scale to get:
$$
\hE_l \sim \Phi, \ \hE_g \sim \ibb_g \Phi^2, \ \hE_{t} \sim (\ibb_g^{1/3} + \ibb_g^{-1/3}).
$$
Though it might seem at first glance that the transition zone energy $\hE_t$ has singular contribution since it diverges in both limits of $\bb_g \rightarrow 0, \infty$, this is however not the case as the transition zone exists only when $\ibb_g, \bb_g \sim \O(1)$. It is also evident from the scaling above that the critical length for transition $\Phi_c \sim \bb_g$ (which is also the case for filaments with free ends as shown in SI sec.~\ref{sec:appdx}).

\subsection*{Coiled phase}
In the coiled phase the packing fraction of the surface of the bubble is independent of $\bb_g$ and is a purely geometric quantity determined by the length, thickness of the filament and the bubble radius. In Fig.~\ref{fig:coils} $(a-d)$ we show the shape of coils in the experiments where we see that the filament after self-contact forms complex shapes but at small bubble radius, aligns with itself, forming multiple coils. When the size of bubble becomes very small, the boundary layer close to the fixed end of the filament determines the overall orientation and this results in the tilt of the coil seen in Fig.~\ref{fig:coils}$(d)$. However in the numerical simulations, in Fig.~\ref{fig:coils}$(e)$, as there is no interaction between filament points, we see that they self-intersect and are all now localized at the bottom of the bubble.

\section*{Conclusion}

\blue{Our results can be posed within the broader framework of \textit{confinement} and \textit{bendability} of elastic structures. Confinement prescribes the geometric constraints placed on the elastic object, which in this case is the constraint of conforming to a spherical substrate of given radius. Bendability determines the competition between forces due to elastic deformation and external body forces, gravity in this case.  Elastic instabilities in thin sheets~\cite{king2012elastic, schroll2013, davidovitch2011prototypical} have been explored within this framework. The experiments presented here show that this paradigm is also applicable to filaments:  the coiling parameter, which depends explicitly on the curvature of the bubble, is a confinement parameter, while the elasto-gravity bendability quantifies the relative magnitude of the forces due to gravity and bending deformation. We have only explored spherical substrates in this work, however, other substrate geometries such as negatively curved surfaces, and surfaces with non-uniform curvature could lead to different instabilities and morphologies.}

\bibliographystyle{unsrtnat}
\bibliography{biblio}

\begin{thebibliography}{26}
\providecommand{\natexlab}[1]{#1}
\providecommand{\url}[1]{\texttt{#1}}
\expandafter\ifx\csname urlstyle\endcsname\relax
  \providecommand{\doi}[1]{doi: #1}\else
  \providecommand{\doi}{doi: \begingroup \urlstyle{rm}\Url}\fi

\bibitem[Huynen et~al.(2016)Huynen, Detournay, and
  Deno{\"e}l]{huynen2016surface}
Alexandre Huynen, Emmanuel Detournay, and Vincent Deno{\"e}l.
\newblock Surface constrained elastic rods with application to the sphere.
\newblock \emph{Journal of Elasticity}, 123\penalty0 (2):\penalty0 203--223,
  2016.

\bibitem[Guven et~al.(2014)Guven, Valencia, and
  V{\'a}zquez-Montejo]{guven2014environmental}
Jemal Guven, Dulce~Mar{\'\i}a Valencia, and Pablo V{\'a}zquez-Montejo.
\newblock Environmental bias and elastic curves on surfaces.
\newblock \emph{Journal of Physics A: Mathematical and Theoretical},
  47\penalty0 (35):\penalty0 355201, 2014.

\bibitem[Swigon et~al.(1998)Swigon, Coleman, and Tobias]{swigon1998elastic}
David Swigon, Bernard~D Coleman, and Irwin Tobias.
\newblock The elastic rod model for dna and its application to the tertiary
  structure of dna minicircles in mononucleosomes.
\newblock \emph{Biophysical journal}, 74\penalty0 (5):\penalty0 2515--2530,
  1998.

\bibitem[Tobias et~al.(2000)Tobias, Swigon, and Coleman]{tobias2000elastic}
Irwin Tobias, David Swigon, and Bernard~D Coleman.
\newblock Elastic stability of dna configurations. i. general theory.
\newblock \emph{Physical Review E}, 61\penalty0 (1):\penalty0 747, 2000.

\bibitem[Coleman et~al.(2000)Coleman, Swigon, and Tobias]{coleman2000elastic}
Bernard~D Coleman, David Swigon, and Irwin Tobias.
\newblock Elastic stability of dna configurations. ii. supercoiled plasmids
  with self-contact.
\newblock \emph{Physical Review E}, 61\penalty0 (1):\penalty0 759, 2000.

\bibitem[Goriely and Neukirch(2006)]{goriely2006mechanics}
Alain Goriely and S{\'e}bastien Neukirch.
\newblock Mechanics of climbing and attachment in twining plants.
\newblock \emph{Physical review letters}, 97\penalty0 (18):\penalty0 184302,
  2006.

\bibitem[Seemann(1996)]{seemann1996deformation}
W~Seemann.
\newblock Deformation of an elastic helix in contact with a rigid cylinder.
\newblock \emph{Archive of Applied Mechanics}, 67\penalty0 (1):\penalty0
  117--139, 1996.

\bibitem[Van~der Heijden(2001)]{van2001static}
GHM Van~der Heijden.
\newblock The static deformation of a twisted elastic rod constrained to lie on
  a cylinder.
\newblock \emph{Proceedings of the Royal Society of London. Series A:
  Mathematical, Physical and Engineering Sciences}, 457\penalty0
  (2007):\penalty0 695--715, 2001.

\bibitem[Tan and Digby(1993)]{tan1993buckling}
XC~Tan and PJ~Digby.
\newblock Buckling of drill string under the action of gravity and axial
  thrust.
\newblock \emph{International journal of solids and structures}, 30\penalty0
  (19):\penalty0 2675--2691, 1993.

\bibitem[Wu et~al.(1993)Wu, Juvkam-Wold, and Lu]{wu1993helical1}
Jiang Wu, HC~Juvkam-Wold, and R~Lu.
\newblock Helical buckling of pipes in extended reach and horizontal wells-part
  1: preventing helical buckling.
\newblock 1993.

\bibitem[Wu and Juvkam-Wold(1993)]{wu1993helical2}
J~Wu and HC~Juvkam-Wold.
\newblock Helical buckling of pipes in extended reach and horizontal wells-part
  2: Frictional drag analysis.
\newblock 1993.

\bibitem[Miller et~al.(2014)Miller, Lazarus, Audoly, and
  Reis]{miller2014shapes}
JT~Miller, Arnaud Lazarus, Basile Audoly, and Pedro~M Reis.
\newblock Shapes of a suspended curly hair.
\newblock \emph{Physical review letters}, 112\penalty0 (6):\penalty0 068103,
  2014.

\bibitem[Mahadevan and Keller(1996)]{mahadevan1996coiling}
L~Mahadevan and Joseph~B Keller.
\newblock Coiling of flexible ropes.
\newblock \emph{Proceedings of the royal society of london. Series A:
  mathematical, Physical and Engineering Sciences}, 452\penalty0
  (1950):\penalty0 1679--1694, 1996.

\bibitem[Mahadevan and Keller(1999)]{mahadevan1999periodic}
L~Mahadevan and Joseph~B Keller.
\newblock Periodic folding of thin sheets.
\newblock \emph{Siam Review}, 41\penalty0 (1):\penalty0 115--131, 1999.

\bibitem[Elettro et~al.(2016)Elettro, Neukirch, Vollrath, and
  Antkowiak]{elettro2016}
Herv{\'e} Elettro, S{\'e}bastien Neukirch, Fritz Vollrath, and Arnaud
  Antkowiak.
\newblock In-drop capillary spooling of spider capture thread inspires hybrid
  fibers with mixed solid--liquid mechanical properties.
\newblock \emph{Proceedings of the National Academy of Sciences}, 113\penalty0
  (22):\penalty0 6143--6147, 2016.

\bibitem[Bico et~al.(2004)Bico, Roman, Moulin, and
  Boudaoud]{bico2004elastocapillary}
Jos{\'e} Bico, Benoit Roman, Loic Moulin, and Arezki Boudaoud.
\newblock Elastocapillary coalescence in wet hair.
\newblock \emph{Nature}, 432\penalty0 (7018):\penalty0 690--690, 2004.

\bibitem[Prasath et~al.(2021)Prasath, Marthelot, Menon, and
  Govindarajan]{prasath2021}
S~Ganga Prasath, Joel Marthelot, Narayanan Menon, and Rama Govindarajan.
\newblock Wetting and wrapping of a floating droplet by a thin elastic
  filament.
\newblock \emph{Soft Matter}, 17\penalty0 (6):\penalty0 1497--1504, 2021.

\bibitem[Brau et~al.()Brau, Ganga~Prasath, and Davidovitch]{brau2018}
Fabian Brau, S~Ganga~Prasath, and Benny Davidovitch.
\newblock Wettability of bendable solids: Insights from a two-dimensional,
  inextensible model.
\newblock \emph{(to be submitted)}.

\bibitem[Schulman et~al.(2017)Schulman, Porat, Charlesworth, Fortais, Salez,
  Rapha{\"e}l, and Dalnoki-Veress]{schulman2017elastocapillary}
Rafael~D Schulman, Amir Porat, Kathleen Charlesworth, Adam Fortais, Thomas
  Salez, Elie Rapha{\"e}l, and Kari Dalnoki-Veress.
\newblock Elastocapillary bending of microfibers around liquid droplets.
\newblock \emph{Soft matter}, 13\penalty0 (4):\penalty0 720--724, 2017.

\bibitem[Py et~al.(2007)Py, Reverdy, Doppler, Bico, Roman, and Baroud]{py2007}
Charlotte Py, Paul Reverdy, Lionel Doppler, Jos{\'e} Bico, Benoit Roman, and
  Charles~N Baroud.
\newblock Capillary origami: spontaneous wrapping of a droplet with an elastic
  sheet.
\newblock \emph{Physical review letters}, 98\penalty0 (15):\penalty0 156103,
  2007.

\bibitem[Audoly and Pomeau(2010)]{audoly2010elasticity}
Basile Audoly and Yves Pomeau.
\newblock \emph{Elasticity and geometry: from hair curls to the non-linear
  response of shells}.
\newblock Oxford university press, 2010.

\bibitem[Kreyszig(2019)]{kreyszig2019differential}
Erwin Kreyszig.
\newblock Differential geometry.
\newblock In \emph{Differential Geometry}. University of Toronto Press, 2019.

\bibitem[Kusumaatmaja and Lipowsky(2011)]{kusumaatmaja2011droplet}
Halim Kusumaatmaja and Reinhard Lipowsky.
\newblock Droplet-induced budding transitions of membranes.
\newblock \emph{Soft Matter}, 7\penalty0 (15):\penalty0 6914--6919, 2011.

\bibitem[King et~al.(2012)King, Schroll, Davidovitch, and
  Menon]{king2012elastic}
Hunter King, Robert~D Schroll, Benny Davidovitch, and Narayanan Menon.
\newblock Elastic sheet on a liquid drop reveals wrinkling and crumpling as
  distinct symmetry-breaking instabilities.
\newblock \emph{Proceedings of the National Academy of Sciences}, 109\penalty0
  (25):\penalty0 9716--9720, 2012.

\bibitem[Schroll et~al.(2013)Schroll, Adda-Bedia, Cerda, Huang, Menon, Russell,
  Toga, Vella, and Davidovitch]{schroll2013}
RD~Schroll, M~Adda-Bedia, E~Cerda, J~Huang, N~Menon, TP~Russell, KB~Toga,
  D~Vella, and B~Davidovitch.
\newblock Capillary deformations of bendable films.
\newblock \emph{Physical review letters}, 111\penalty0 (1):\penalty0 014301,
  2013.

\bibitem[Davidovitch et~al.(2011)Davidovitch, Schroll, Vella, Adda-Bedia, and
  Cerda]{davidovitch2011prototypical}
Benny Davidovitch, Robert~D Schroll, Dominic Vella, Mokhtar Adda-Bedia, and
  Enrique~A Cerda.
\newblock Prototypical model for tensional wrinkling in thin sheets.
\newblock \emph{Proceedings of the National Academy of Sciences}, 108\penalty0
  (45):\penalty0 18227--18232, 2011.

\end{thebibliography}

	\large
	\widetext
	\begin{center}
		\textbf{\large Supplemental Materials: Shapes of a filament on the surface of a bubble}
	\end{center}
\setcounter{equation}{0}
\setcounter{figure}{0}
\setcounter{table}{0}
\setcounter{page}{1}
\makeatletter

\renewcommand{\thetable}{S\arabic{table}}%
\renewcommand{\thesection}{S\arabic{section}}
\renewcommand{\theequation}{S\arabic{equation}}
\renewcommand{\thefigure}{S\arabic{figure}}
\renewcommand{\theHtable}{Supplement.\thetable}
\renewcommand{\theHfigure}{Supplement.\thefigure}

\section{Coiling with two free ends}\label{sec:appdx}
\begin{figure}
	\centering
	\includegraphics[width=0.75\textwidth]{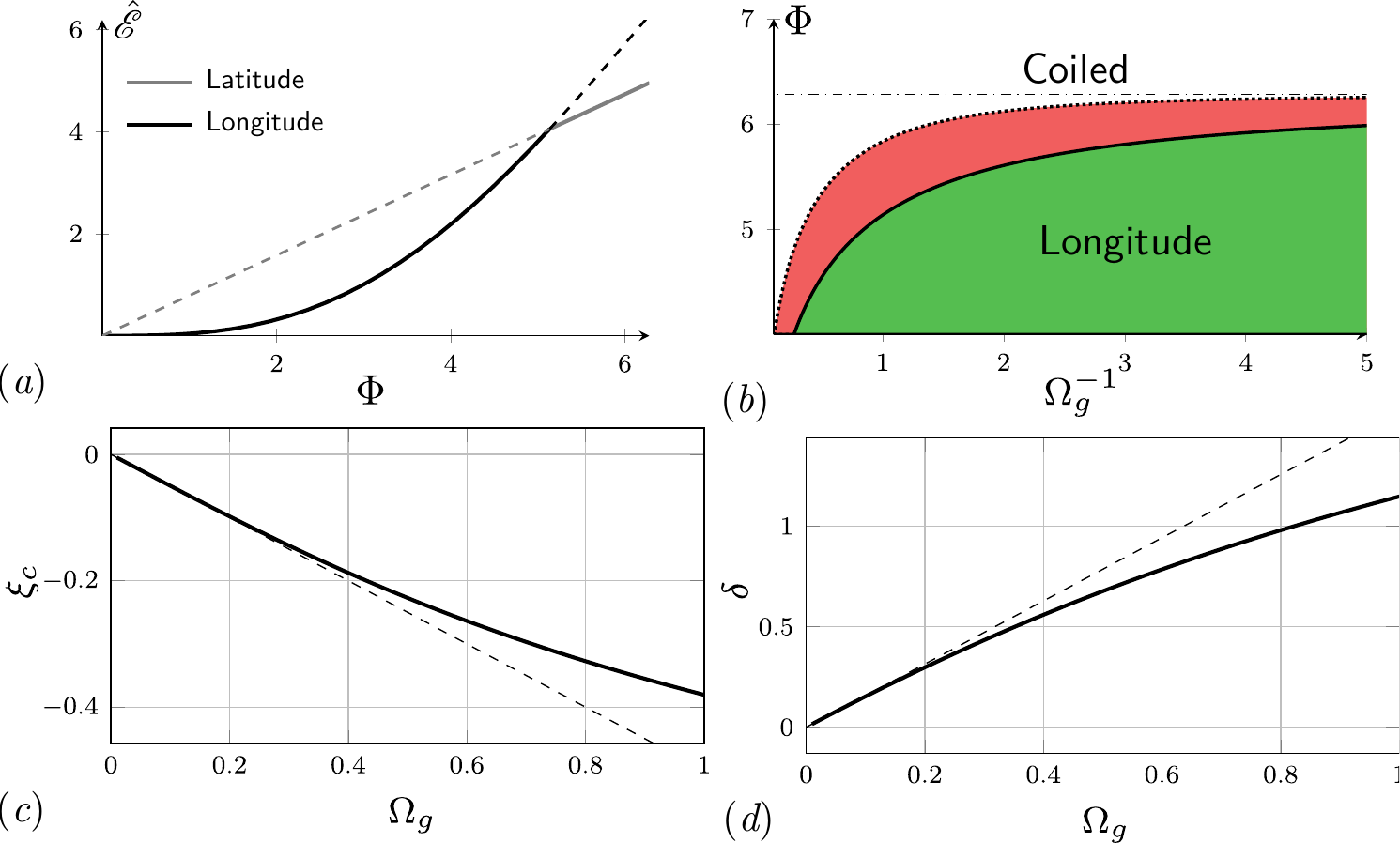}
	\caption{$(a)$ Total energy of a filament as a function of its non-dimensional length, $\varphi$ when it lies on a longitude and a latitude for $\bb_g=1$. Filaments choose the minimum of these and we find the mode switch happen at $\varphi_c=5.13$. $(b)$ Phase boundary defined based on the intersection of energies $\hE_l, \hE_g$ in Eq.~\ref{eq:equator},~\ref{eq:long}. The filament switches modes beyond $\Phi_c$ to the minimum of the energies, indicated here as boundary of the green region with a solid line.}
	\label{fig:SIsuppl}
\end{figure}

In our experiments we held one end of the filament fixed at the bubble's north pole. However when both the ends are free the calculation of the critical length $\Phi_c$ for transition between different morphologies gets simplified. The filament shape is given by either the geodesic or the polar angle of the latitude based on the energy, which is a function of the coiling parameter and elasto-gravity bendability. In order to calculate the minimum energy, we use the expressions in Eq.~\ref{eq:equator},~\ref{eq:long} (see Fig.~\ref{fig:SIsuppl}$(a)$ for the functional form when $\bb_g=1$). The minimum energy stays along the geodesic up to a critical value of $\Phi\le 5.13$ when $\bb_g=1$. Beyond this critical value of coiling parameter the solution branch shifts to that of a latitude (shown by a solid gray line), as the energy of the geodesic exceeds that of the latitude of the same length. The solid line in Fig.~\ref{fig:SIsuppl}$(b)$ is the critical coiling parameter $\varphi_c$ when the filament transitions from a geodesic to a different latitude calculated as a function of $\Phi$ for different values of $\bb_g$. The filament starts coiling only when the filament wraps a given latitude completely and comes into self-contact. This is indicated by the dashed line in Fig.~\ref{fig:SIsuppl}$(b)$, which is calculated by using the relation $\varphi^* = 2\pi \cos(\xi_c)$, where $\xi_c$ is the polar angle of the latitude configuration. 

\subsection*{Asymptotic behaviour}
When both ends of the filament are free, we expect $\Phi_c \rightarrow 2\pi$ in the limit of zero gravity just as seen in the main text. Thus, when $\ibb_g \rightarrow 0$ the selected latitude is described by polar angle $\xi_c \rightarrow 0$. We have two small parameters $\xi_c$, $\ibb_g$ whose relationship is evaluated by expanding the solution close to $\Phi \rightarrow 2\pi$ as $\ibb_g \rightarrow 0$. From the minimised solution we know that the latitude's polar angle satisfies,
\[
\ibb_g \cos^4 \xi_c = -2 \sin \xi_c.
\]
Expanding the above expression, we get the leading order behaviour:
\[
\xi_c \sim -\frac{\ibb_g}{2}.
\]
We find that $\xi_c$ is independent of $\Phi$ after the transition from the geodesic i.e. $\Phi > \Phi_c$. Further as gravity becomes stronger the polar angle to which filament migrates moves towards south pole. This comes from a compromise between bending energy and potential energy where filament chooses to bend more to reduce the potential energy with increase in $\ibb_g$. In order to find the relationship between the critical filament length $\Phi_c$ at which bifurcation happens in this asymptotic limit, we expand the full energy expression for a latitude in Eq.~\ref{eq:equator} to get,
\begin{align}
	\hE_g &= \Phi(\ibb_g (\sin \xi_c + z_o) + 1 + \tan^2(\xi_c)), \\
	&\approx \Phi \bigg(1 - \frac{\ibb_g^2}{4} \bigg).
\end{align}
A similar expansion for the energy of a geodesic can be performed to get
\begin{align}
	\hE_l &= \Phi - 2 \ibb_g \sin\frac{\Phi}{2}.
\end{align}
Since we are interested in the region close to $\Phi \rightarrow 2\pi$, we have the small parameter $\delta=(2\pi - \Phi)$. The geodesic energy can be written in terms of $\delta$ as:
\begin{align}
	\hE_l &\approx 2\pi - \delta - \ibb_g \delta + \mathcal{O}(\delta^3)
\end{align}
The criteria for transition is found by solving $\hE_g = \hE_l$ and this gives,
\begin{align}
	\frac{\ibb_g^2 \delta}{4}-\frac{\pi \ibb_g^2}{2}&= -\ibb_g\delta, \\
	\delta &\approx \frac{\pi \ibb_g}{2}.
\end{align}
We plot in Fig.~\ref{fig:SIsuppl}$(c,d)$ the asymptotic expressions derived above for $\xi_c$ and $\delta$ as dashed lines and compare it with the full solution shown as solid curves.

\section{Regularizing singularities}\label{sec:reg}
In Eq.~\ref{eq:enerFix} we have singularities at the south pole of the bubble because the curvature of the fraction $(1-\alpha)$ of the filament along the latitude diverges as the square of curvature of the latitude portion, whereas the fraction of length along latitude $(1-\alpha) \rightarrow 0$ contributes only linearly. In order to avoid this divergence, we need a regularization length scale arising from the missing length-scale in the model, the bendo-capillary length $\lec$. To resolve this divergence we multiply the singular contribution with a function $\Lambda (\beta)$ that suppresses the singularity. The energy then becomes,
\begin{align}
\hE =& \ \alpha \Phi + \ibb_g \sin(\alpha \Phi) + \ibb_g \Phi (1 - \alpha) \cos(\alpha \Phi) \nonumber \\
& \ + (1 - \alpha) \Phi \big\{ [1 - \Lambda (0) - \Lambda (\pi) - \Lambda (2\pi) ]\csc^2(\alpha \Phi) \big\}, 
\end{align}
\[
\text{where} \ \Lambda (\beta) = \ \exp \bigg(-\frac{(\alpha \Phi -\beta)^2}{\eta(\ibb)} \bigg).
\]
Here $\eta(\ibb)$ is the cut-off non-dimensional length-scale associated with the capillary bendability, as this is the length scale that acts at the boundary of scale separation. However owing to the limits of scale-separation when the radius of curvature along a latitude approaches the length-scale associated with capillary bendability, this energy expression is no more valid. These regularizing terms are relevant at three locations along the filament i.e. $\alpha \Phi = 0, \pi, 2\pi$, as all these locations result in diverging values of the energy.

\end{document}